\documentclass[prc,aps,floatfix,twocolumn,groupedaddress,nofootinbib,showpacs,preprintnumbers,
amsmath,amssymb,amsfonts,widetable] {revtex4-1}
\usepackage{bm}
\usepackage{soul}
\usepackage{array}
\usepackage{lipsum}
\usepackage{relsize}
\usepackage{mathrsfs}
\usepackage{amssymb}
\usepackage{amsmath}
\usepackage{graphicx}
\usepackage{slashed}
\usepackage[usenames]{color}
\usepackage{pslatex}
\usepackage{booktabs}
\usepackage{physics}

\begin{document}
\title{Building an Equation of State Density Ladder} 

\author{Marc Salinas and J. Piekarewicz}

\affiliation{Department of Physics, Florida State University, 
Tallahassee, FL 32306, USA}

\date{\today}

\begin{abstract}
The confluence of major theoretical, experimental, and observational advances
are providing a unique perspective on the equation of state of dense neutron-rich 
matter---particularly its symmetry energy---and its imprint on the 
mass-radius relation for neutron stars. In this contribution we organize these 
developments in an equation of state density ladder. Of particular relevance 
to this discussion is the impact of the various rungs on the equation of state 
and the identification of possible discrepancies among the various methods. 
A preliminary analysis identifies a possible tension between laboratory 
measurements and gravitational-wave detections that could indicate the 
emergence of a phase transition in the stellar core.
\end{abstract}

\maketitle

\section{Introduction}

\emph{What are the new states of matter that emerge at exceedingly high density and 
temperature?} featured among one of the ``Eleven Science Questions for the New 
Century" posed by the Committee on the Physics of the Universe at the turn of the past 
century\,\cite{QuarksCosmos:2003}. Closely related to this question and one at the 
center of nuclear science today is, \emph{How does subatomic matter organize itself and 
what phenomena emerge?}\,\cite{LongRangePlan}. Isolated neutron stars are ideal cosmic 
laboratories to study the emergence of new states of matter over an enormous range of 
densities spanning more than 10 orders of magnitude. 

Neutron stars are cold, fully catalyzed astronomical objects that settle into the absolute 
ground state at the appropriate baryon density. In this manner, neutron stars provide a
unique laboratory where a density can be dialed and the ground state of the system at
such density explored. Moreover, being bound by gravity and not by the strong force, 
conditions found in the neutron-star interior are impossible to reproduce in terrestrial
laboratories.

\vspace{5pt}
\begin{center}
\begin{figure}[h]
\centering
\includegraphics[width=0.45\textwidth]{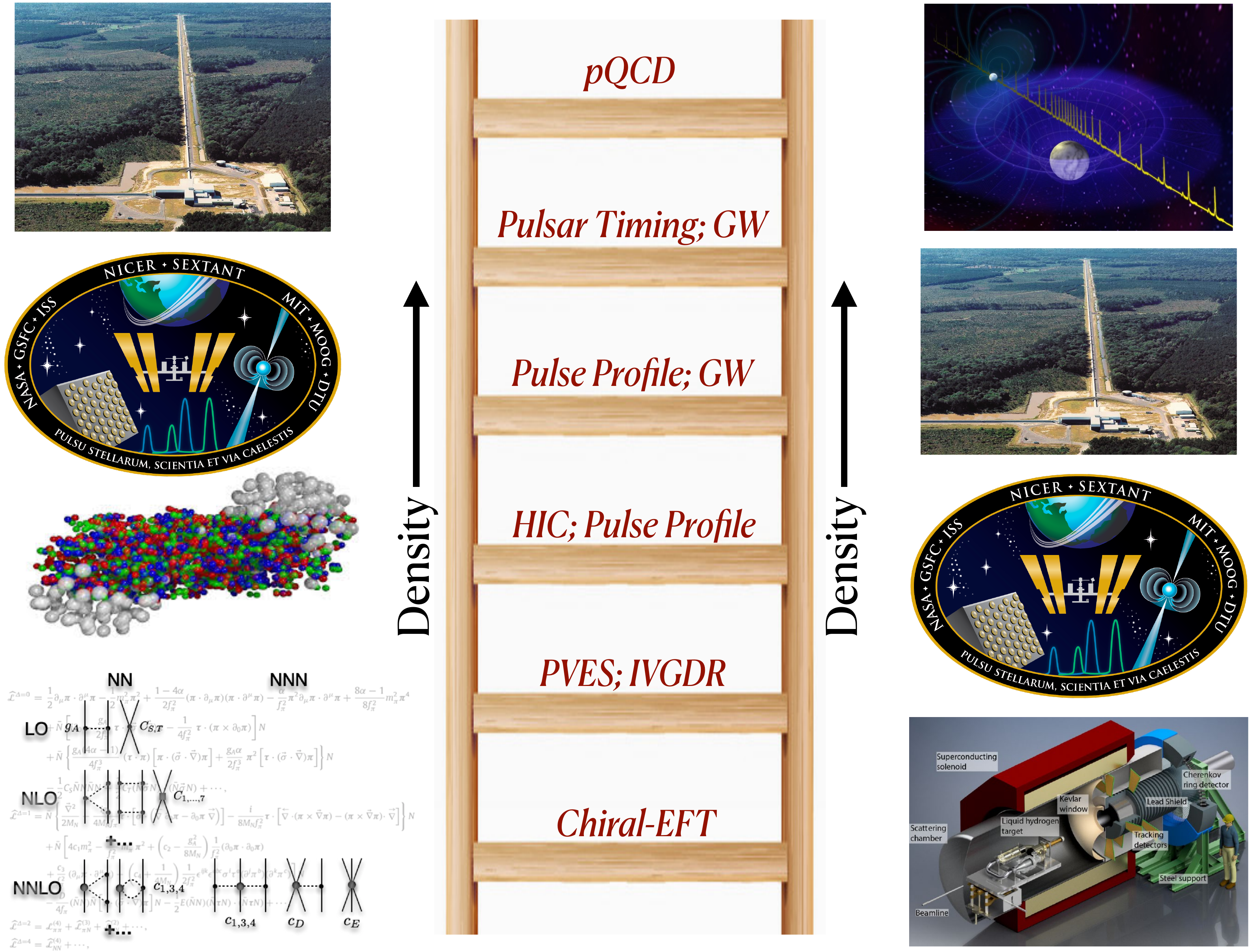}
\caption{The equation of state density ladder. Each rung in the ladder represents 
a technique that informs the equation of state of neutron rich matter in a suitable 
density regime. With the exception of the isovector giant dipole resonance (IVGDR), 
Heavy Ion Collisions (HIC), and perturbative QCD (pQCD), the impact of all other 
methods on the EOS have been addressed in this contribution.}
\label{Fig1}
\end{figure}
\end{center}

Powerful insights into the structure, dynamics, and composition of neutron stars have 
emerged during the last few years from discoveries that probe different regions of the 
stellar interior. Given the confluence of such discoveries, it is fitting to introduce an 
``equation of state density ladder"\,\cite{Piekarewicz:2022ycz}, akin to the cosmic 
distance ladder used in cosmology, to illustrate the different techniques that are being 
used to probe the various regions of the neutron star. In this manner each rung in the 
ladder represents a theoretical, experimental, or observational technique that determines 
the equation of state (EOS) in a suitable density regime. Paraphrasing from the cosmic 
distance ladder, no one method can determine the EOS over the entire density domain 
existent in a neutron star. Rather, each rung on the ladder informs the EOS in a density 
regime that can be connected to its neighboring rungs. Such a density ladder has been 
constructed in Fig.\ref{Fig1} using a variety of theoretical, experimental, and observational 
methods that have propelled the field to a golden era of neutron 
stars\,\cite{Baym:2019,Baym:2017whm}. In the next few sections we will describe the 
remarkable progress made in these various arenas and examine their profound implication 
in constraining the equation of state of dense matter.

\section{Chiral Effective Field Theory}
\label{ChEFT}

The lowest rung in the density ladder displayed in Fig.\ref{Fig1} is a purely theoretical
approach developed by Steven Weinberg in the early 1990s. In a paper entitled 
``Nuclear forces from chiral lagrangians''\,\cite{Weinberg:1990rz}, Weinberg incorporated 
the approximate chiral symmetry of Quantum Chromodynamics into the construction of 
a nuclear Hamiltonian, where the long-range part of the interaction is mediated by the pion 
and the (unknown) short-range structure is encoded in contact interactions with empirical
coefficients fitted to the data. Such a seminal paper has triggered a paradigm shift in
theoretical nuclear physics. In essence, chiral effective field theory ($\chi$EFT) is a 
systematic, improvable, and quantifiable theoretical framework of nucleons interacting 
via the exchange of pions and unresolved short-range structure encoded in a few 
contact terms. Whereas $\chi$EFT combined with a variety of many-body methods 
have significantly advanced our understanding of the atomic nucleus---see for example 
\,\cite{Bedaque:2002mn,Epelbaum:2008ga,Machleidt:2011zz,RodriguezEntem:2020jgp}
and references contained therein---in the present contribution we focus on the impact 
$\chi$EFT has on our understanding of dense matter.

Given that nuclear matter saturates, namely, there is an equilibrium density of about
$\rho_{0}\!\approx\!0.15\,{\rm fm}^{-3}$ that characterizes the interior density of medium 
to heavy nuclei, probing the nuclear dynamics below this density is particularly challenging.  
Chiral effective field theory offers the only realistic method to constrain the EOS below 
saturation density. Indeed, $\chi$EFT predictions of increasing quality and sophistication
are providing valuable insights on the EOS of pure neutron matter for densities below
$1.5\rho_{0}$\,\cite{Hebeler:2009iv,Tews:2012fj,Kruger:2013kua,
Lonardoni:2019ypg,Drischler:2020hwi,Drischler:2021kxf,Sammarruca:2021mhv,
Sammarruca:2022ser}. We note that as in the case of all effective field theories, 
$\chi$EFT includes a break-down scale that defines the range of applicability of the 
theory. In the case of infinite nuclear matter, the approach is valid provided the Fermi 
momentum is below the break-down scale. It is precisely in this sense that no one 
method depicted in the density ladder can determine the EOS over the entire density domain. 

Two quantities that will be used throughout this paper to compare the various 
approaches are the slope of the symmetry energy at saturation density ($L$) and the 
radius of a $1.4\,M_{\odot}$ neutron star. The symmetry energy quantifies the energy 
cost in turning symmetric nuclear matter into pure neutron matter and is defined as 
follows\,\cite{Piekarewicz:2008nh}:
\begin{equation}
 S(\rho) = \frac{1}{2} \left(\frac{\partial^{2}\varepsilon(\rho,\alpha)}
               {\partial\alpha^{2}}\right)_{\!\!\alpha=0},
\label{SymmE0}
\end{equation}
where $\varepsilon(\rho,\alpha)$ is the energy per nucleon of infinite nuclear matter
that depends on the sum and difference of proton and neutron densities; that is,
$\rho\!\equiv\!\rho_{p}\!+\!\rho_{n}$ and
$\alpha\!\equiv\!(\rho_{p}\!-\!\rho_{n})/\rho$ respectively. Note that infinite nuclear matter is an
idealized system of neutrons and protons interacting exclusively via the strong force,
without contribution from the electromagnetic or weak interactions. Moreover, it is often
convenient to characterize the behavior of the symmetry energy around saturation
density in terms of a few bulk parameters, namely\,\cite{Piekarewicz:2008nh},
\begin{equation}
 S(\rho) = J + L\,x + \frac{1}{2} K_{\rm sym}\,x^{2} + \ldots 
 \quad x\!\equiv\!\frac{(\rho-\rho_{0})}{3\rho_{0}},
\label{SymmE1}
\end{equation}
where $J$, $L$, and $K_{\rm sym}$ are the values of the symmetry energy, its slope, 
and curvature at saturation density. Given that the slope of the symmetry energy is
closely related to the pressure of pure neutron matter at saturation density, namely, 
\begin{equation}
 P_{0}=\frac{1}{3}\rho_{0}L,
 \label{PvsL}
\end{equation}
this quantity is of critical importance as it determines both the neutron skin of heavy 
nuclei\,\cite{Brown:2000,Furnstahl:2001un,RocaMaza:2011pm,Piekarewicz:2019ahf} 
as well as the radius of low-mass neutron 
stars\,\cite{Horowitz:2000xj,Horowitz:2001ya,Carriere:2002bx}. In particular, $\chi$EFT 
predicts a value for the slope of the symmetry energy of
$L\!=\!(59.8\!\pm\!4.1)\,{\rm MeV}$ at the $1\sigma$ level\,\cite{Drischler:2020hwi}.

\section{Parity Violating Electron Scattering: the neutron skin thickness of 
${}^{208}$P\lowercase{b}}

More than three decades ago, Donnelly, Dubach, and Sick proposed the use of 
Parity Violating Electron Scattering (PVES) as a clean and model-independent 
probe of neutron densities\,\cite{Donnelly:1989qs}. The interest in measuring
the neutron distribution of heavy nuclei (specifically of ${}^{208}$Pb) was 
rekindled because of the enormously successful experimental program developed 
at the Thomas Jefferson National Accelerator Facility (Jefferson Lab) and by the 
impact that such a measurement could have in constraining the equation of state 
of neutron rich matter and ultimately the structure of neutron 
stars\,\cite{Horowitz:2000xj,Horowitz:2001ya}.

The parity-violating asymmetry is defined as the difference relative to the
sum of the differential cross section for the elastic scattering of right/left-handed 
longitudinally polarized electrons. In a simple plane-wave impulse approximation,
the asymmetry emerges from the interference between two Feynman diagrams,
a large one involving the exchange of a photon and a much smaller one involving 
the exchange of a $Z^{0}$ boson. As such, the parity violating asymmetry becomes
\begin{align}
   A_{PV}(Q^{2}) &= \frac{\displaystyle{\left(\frac{d\sigma}{d\Omega}\right)_{\!\!R}  - 
                                            \left(\frac{d\sigma}{d\Omega}\right)_{\!\!L}}}
                        {\displaystyle{\left(\frac{d\sigma}{d\Omega}\right)_{\!\!R}  + 
                        \left(\frac{d\sigma}{d\Omega}\right)_{\!\!L}}}, \nonumber \\
                 &= \frac{G_{\!F}Q^{2}}{4\pi\alpha\sqrt{2}}
                              \frac{Q_{\rm wk}F_{\rm wk}(Q^{2})}{ZF_{ch}(Q^{2})}.
\label{APVb}
\end{align}
where $Q^{2}$ is the square of four-momentum transfer to the nucleus, $\alpha$ is
the fine-structure constant, and $G_{F}$ is the Fermi constant. In turn, the nuclear
information is contained in the electric charge of the nucleus $Z$, its weak-vector
charge $Q_{\rm wk}\!=\!-N\!+\!(1\!-\!4\sin^{2}\!\theta_{\rm W})Z$, and two
form factors $F_{\rm wk}$ and $F_{ch}$, both normalized to one at $Q^{2}\!=\!0$. 
First, we note that because the weak charge of the proton is small, most of the weak 
charge of the nucleus is carried by the neutrons. Second, the two nuclear form factors 
are proportional to the Fourier transform of their respective densities. Finally, given 
that the charge form factor for a great number of nuclei is known with enormous 
precision\,\cite{Angeli:2013}, the one remaining unknown in the problem is $F_{\rm wk}$. 
And because the weak charge of the nucleus resides largely on the neutrons, the parity 
violating asymmetry---as first suggested in Ref.\,\cite{Donnelly:1989qs}---provides an 
ideal, model-independent experimental tool to determine neutron densities.

Although it took decades since first suggested by Donnelly, Dubach, and Sick, the Lead 
Radius EXperiment (PREX) at Jefferson Lab fulfilled its promise to determine the neutron 
radius of ${}^{208}$Pb with a precision of nearly 1\%\,\cite{Abrahamyan:2012gp,Horowitz:2012tj,
Adhikari:2021phr}. In particular, the neutron skin thickness of ${}^{208}$Pb---defined as the 
difference between the neutron and proton root-mean-square radii------was reported at the 
$1\sigma$ level to be\,\cite{Adhikari:2021phr}:
\begin{equation}
 R_{\rm skin}=R_{n}-R_{p}=(0.283\pm0.071)\,{\rm fm},
 \label{Rskin}
\end {equation} 
Although the error is large, the central value is much larger than previously anticipated both by 
previous experimental and theoretical estimates\,\cite{Thiel:2019tkm,Reed:2021nqk}. Indeed,
by relying on the strong correlation between the neutron skin thickness of ${}^{208}$Pb and 
the slope of the symmetry energy, a value of $L\!=\!(106\pm37)\,{\rm MeV}$ was 
obtained\,\cite{Thiel:2019tkm,Reed:2021nqk}. In turn, the large value of $L$ implies a 
correspondingly large value for the radius of a $1.4\,M_{\odot}$ neutron star of 
$13.25\!\lesssim\!R_{1.4}({\rm km})\!\lesssim\!14.26$, suggesting that the symmetry 
energy is fairly stiff\,\cite{Reed:2021nqk}. It is worth noting that the extraction of $L$ from the PREX 
experiment is significantly larger than the $\chi$EFT prediction of $L\!=\!(59.8\!\pm\!4.1)\,{\rm MeV}$. 

It has been argued that such a discrepancy may just be a statistical fluctuation, given that the 
two values agree at the $2\sigma$ level. At present, the only possibility of resolving whether 
the tension is real, is at the future Mainz Energy-recovery Superconducting Accelerator (MESA) 
being under construction at the Johannes Gutenberg University in Mainz, Germany\,\cite{Becker:2018ggl}.
If the Mainz Radius EXperiment (MREX) becomes feasible, one can anticipate a factor-of-two 
improvement in the determination of the neutron radius of ${}^{208}$Pb relative to PREX.

\section{LIGO-Virgo: neutron star mergers}
\label{LVC}

The historic detection of gravitational waves emitted from the binary neutron star merger 
GW170817 is providing fundamental new insights into the nature of dense 
matter\,\cite{Abbott:PRL2017}. Of great relevance to the equation of state are the so-called 
``chirp mass'' and ``chirp tidal deformability'' (or simply $\tilde{\Lambda}$) given respectively 
by 
\begin{subequations}
\begin{align}
 & \hspace{2cm} {\cal M} = \frac{(M_{1}M_{2})^{3/5}}{(M_{1}+M_{2})^{1/5}}, \\
 & \tilde{\Lambda} = \frac{16}{13}
 \frac{(M_{1}+12M_{2})M_{1}^{4}\Lambda_{1} +
         (M_{2}+12M_{2})M_{2}^{4}\Lambda_{2}}
         {(M_{1}+M_{2})^{5}},
\end{align}
\end{subequations}
where the dimensionless tidal deformability of an individual neutron star of mass $M$ 
and radius $R$ is defined as\,\cite{Hinderer:2007mb,Hinderer:2009ca,Damour:2009vw,
Postnikov:2010yn,Fattoyev:2012uu,Steiner:2014pda,Fattoyev:2017jql}.
\begin{equation}
 \Lambda = \frac{2}{3}k_{2}\left(\frac{c^{2}R}{GM}\right)^{\!5}
                 = \frac{64}{3}k_{2}\left(\frac{R}{R_{s}}\right)^{\!5}.
 \label{Lambda}
\end{equation}
Here $k_{2}$ is the second Love number that is mildly sensitive to the equation of
state and $R_{s}$ is the Schwarzschild radius of the neutron star. Note that for the 
equal mass case, $\tilde{\Lambda}\!=\!\Lambda_{1}\!=\!\Lambda_{2}$. The tidal 
deformability is extremely sensitive to the equation of state as it scales as the fifth 
power of the compactness parameter $M/R$. The tidal 
field of the companion star induces a mass quadrupole moment in the neutron star 
that---in  the linear regime---is proportional to the tidal field; the constant of proportionality 
is the tidal deformability. Thus, for a given mass, a larger (more ``fluffy") neutron star 
is easier to tidally deform than a corresponding smaller star. 

Whereas the chirp mass of GW170817 was determined with enormous precision 
(about a few parts in a thousand) the tidal deformability hides behind the fifth 
post-Newtonian coefficient in the waveform. So at the time of the discovery paper, only an upper 
bound on $\tilde{\Lambda}$ was reported\,\cite{Abbott:PRL2017}. Yet in a follow-up 
paper\,\cite{Abbott:2018exr}, the LIGO-Virgo collaboration was able to quote a value 
for the dimensionless tidal deformability of a $1.4\,M_{\odot}$ neutron star of
$\Lambda_{1.4}\!=\! 190^{+390}_{-120}$, favoring soft EOSs, namely, those equations 
of state for which the pressure increases slowly with increasing density. In turn, soft
equations of state predict compact stars with relatively small stellar radii. 

\section{NICER: simultaneous determination of masses and radii of neutron stars}
\label{NICER}

Besides the tidal deformability, electromagnetic emissions from stellar hot spots are also 
highly sensitive to the compactness parameter. The Neutron Star Interior Composition 
Explorer (NICER) monitors soft X-rays emitted from the stellar hot spots by relying on 
the powerful technique of Pulse Profile Modeling\,\cite{Psaltis:2013fha,Watts:2016uzu}. 
As the neutron star spins, traditional Newtonian gravity predicts an oscillating profile with 
no electromagnetic detection once the hot spots move away from the line of sight. 
However, one of the hallmarks of general relativity is gravitational light bending. This
implies that the X-ray emissions, while modulated may never completely disappear;
NICER can ``see" the back of the star. As gravitational light bending increases with
increasing compactness $M/R$, a precise determination of the X-ray profile provides
critical information on the EOS. 

Remarkably, prior to the deployment of NICER in 2017 no single neutron star had both 
their mass and radius simultaneously determined, even though the first pulsar was 
detected by Jocelyn Bell back in 1967\,\cite{Hewish:1968}. Since then, NICER has
reported simultaneous determinations of masses and radii for two neutron stars. The
first mass-radius determination focused on the millisecond pulsar PSR J0030+0451,
with a mass in the neighborhood of the ``canonical" mass of about 
$1.4\,M_{\odot}$\,\cite{Riley:2019yda,Miller:2019cac}. The second target of the NICER 
mission was the millisecond pulsar PSR J0740+6620\,\cite{Riley:2021pdl,Miller:2021qha}. 
Although fainter than PSR J0030+0451, the great advantage of PSR J0740+6620 was 
that its mass was already known. Indeed, as discussed in Sec.\,\ref{PTiming}, with a mass 
in excess of two solar masses, PSR J0740+6620 is currently the heaviest well-measured 
neutron star\,\cite{Cromartie:2019kug,Fonseca:2021wxt}. It is interesting to note that the
stellar radii of both PSR J0030+0451 and PSR J0740+6620 are very close to each other;
about 12.4\,km. This result seems to validate a conjecture that suggest that neutron stars 
have approximately the same radius over a wide range of masses\,\cite{Guillot:2013wu}.
Moreover, that the radius is relatively large implies---unlike GW170817---that the equation 
of state is relatively stiff. Whereas this may indicate a mild tension, the error bars are
currently too large to make a definite statement. 

\section{Pulsar Timing: determination of the most massive neutron stars}
\label{PTiming}

The most stringent constraints on the high density component of the EOS are placed by
the most massive neutron stars. Unlike stellar radii that are sensitive to the EOS in the
vicinity of twice saturation density, massive neutron stars inform the EOS at the highest
densities achieved in the core. In particular, PSR J0740+6620 with a mass of 
$M\!=\!2.08\pm0.07\,M_{\odot}$ has, until recently, been identified as the most massive 
neutron star to date\,\cite{Cromartie:2019kug,Fonseca:2021wxt}. The massive pulsar was 
detected by the Green Bank Telescope using Shapiro delay\,\cite{Shapiro:1964}, often 
regarded as the fourth test of general relativity. The main concept behind Shapiro delay 
is that as the electromagnetic radiation emitted by the neutron star experiences a time
delay as it ``dips" into the gravitational well induced by it white-dwarf companion on its way 
to the detector; no such delay exists when the neutron star is between the white dwarf and 
the observer. By accounting for every orbital period over long periods of time, pulsar timing 
provides a highly precise value for the mass of the white-dwarf star. Now using Kepler's
third law of planetary motion, which is only sensitive to the sum of the individual masses,
one can then extract the mass of PSR J0740+6620.

The record for the most massive neutron star was broken last year with the measurement 
of the mass of the black widow pulsar PSR J0952-0607. As part of a binary system with a
faint sub-solar mass companion, the mass of PSR J0952-0607 was determined to be
$M\!=\!2.35\pm0.17\,M_{\odot}$\,\cite{Romani:2022jhd}, a value that is likely to be near 
the upper limit for non-rotating neutron stars. Indeed, an analysis of GW190814---a
gravitational wave detection from the coalescence of a 23\,$M_{\odot}$ black hole with 
a 2.6\,$M_{\odot}$ compact object, seems to suggest that GW190814 is unlikely to 
originate in a neutron star-black hole coalescence\,\cite{Abbott:2020khf}. Such a claim
is validated by an analysis of the ejecta during the spin-down phase of GW170817
which places an upper limit on the maximum neutron star mass at
${M}_{\max }\lesssim 2.17\,{M}_{\odot }$\,\cite{Margalit:2017dij}. Regardless of the precise 
value of the maximum neutron star mass, it is clear that the EOS at the highest densities 
found in the stellar core must be stiff.

\section{Results}
\label{Results}

In this section we collect the theoretical, experimental, and observational information 
presented in the previous sections, and depicted on the EOS density ladder of 
Fig.\ref{Fig1}, to discuss the new set of energy density functionals introduced in 
Ref.\cite{Salinas:2023nci} and its impact on the mass-radius relation.

To describe the nuclear dynamics and calculate ground states properties of finite nuclei we employ 
the following effective interacting Lagrangian\,\cite{Walecka:1974qa,
Boguta:1977xi,Serot:1984ey,Mueller:1996pm,Serot:1997xg,Horowitz:2000xj}:
\begin{equation}
\begin{split}
{\mathscr L}_{\rm int} =&
\bar\psi \left[g_{\rm s}\phi - \left(g_{\rm v}V_\mu + \frac{g_{\rho}}{2}{\mbox{\boldmath $\tau$}}\cdot{\bf b}_{\mu} + \frac{e}{2}(1 + \tau_{3})A_{\mu}\right)\gamma^{\mu} \right]\psi \\ 
&- \frac{\kappa}{3!} (g_{\rm s}\phi)^3 - \frac{\lambda}{4!}(g_{\rm s}\phi)^4 + \frac{\zeta}{4!} g_{\rm v}^4(V_{\mu}V^\mu)^2 \\
&+ \Lambda_{\rm v}\Big(g_{\rho}^{2}\,{\bf b}_{\mu}\cdot{\bf b}^{\mu}\Big) \Big(g_{\rm v}^{2}V_{\nu}V^{\nu}\Big)\;,
\end{split}
\label{lagrangian}
\end{equation}
where $\psi$ is the isodoublet nucleon field, $A_{\mu}$ is the photon field, and 
$\phi$, $V_{\mu}$, and ${\bf b}_{\mu}$ represent the isoscalar-scalar 
$\sigma$-meson, the isoscalar-vector $\omega$-meson, and the isovector-vector 
$\rho$-meson fields, respectively. The $\sigma$-meson is responsible for the 
intermediate range attraction of the nuclear force, the $\omega$-meson mediates 
the repulsion at short distances, while the $\rho$-meson induces an isospin
dependence that significantly impacts the nuclear symmetry energy. 

For spatially non-uniform systems, we use the Kohn-Sham equations developed in the
framework of density functional theory\,\cite{Hohenberg:1964zz,Kohn:1965}, which
closely resemble a mean-field like approach. For spherically symmetric nuclei, we 
solve the resulting sets of meson and nucleon field equations self consistently;
see\,\cite{Giuliani:2022yna} and references contained therein. For a given nucleus, 
the observables obtained from such a self-consistent procedure are the binding energy 
per nucleon, the single-particle energies and associated Dirac orbitals, the resulting
meson fields, and proton and neutron densities. In turn, by appropriately folding the
proton and neutron densities with single-nucleon form factors determined experimentally,
one can predict charge and weak-charge densities\,\cite{Horowitz:2012we}. Particularly
relevant to this work are proton, neutron, charge, and weak charge radii. In
the particular case of the charge radius, this is obtained as follows:
\begin{equation}
R_{\rm ch}^2 = \frac{1}{Z}\int r^2 \rho_{\rm ch}(r) d^{3}r
=\frac{4 \pi}{Z} \int_0^\infty r^4 \rho_{\rm ch}(r) dr,
\label{rms_charge}
\end{equation}
where we have used the spherical symmetry of the ground state densities. Similar
expressions may be written for the proton, neutron, and weak charge radii. Finally,
given their importance in constraining the slope of the symmetry energy $L$, we 
define neutron skins and weak skins\,\cite{Horowitz:2012we} by 
\begin{subequations}
 \begin{align}
  R_{\rm nskin} & = R_{n}-R_{p} \\
  R_{\rm wskin} & = R_{wk}-R_{ch}.    
 \end{align}
\label{skins}
\end{subequations}

For a given set of coupling constants and meson masses, $\textbf{C} = \{m_{\rm s}, m_{\rm v}, m_{\rho}, g_{\rm s}, 
g_{\rm v}, g_{\rho}, \kappa, \lambda, \zeta, \Lambda_{\rm v} \}$, which we treat as the parameters of the model, 
we can obtain both properties of finite nuclei and the EOS for neutron star matter (NSM). Table \ref{Table1} 
contains the four different sets of parameters that are used in this paper. The first two sets, FSUGold2 and 
FSUGarnet, were calibrated to properties of finite nuclei such as binding energies, charge radii, and giant 
monopole resonances (GMR) \cite{Chen:2014sca,Chen:2014mza}. The last two sets of parameters denote a recently 
refined version of the two models mentioned above. In this ``re-calibration" of the models, input from $\chi$EFT, 
mass-radius measurements from NICER, and tidal deformability from LIGO-Virgo, were used in a Bayesian framework to provide updated sets of coupling constants. More information on this method can be found in\,\cite{Salinas:2023nci}.

\onecolumngrid
\begin{center}
\begin{table}[h]
\begin{tabular}{|l||c|c|c|c|c|c|c|c|c|c|}
\hline\rule{0pt}{2.5ex}   
\!Model   &  $m_{\rm s}$  &  $m_{\rm v}$  &  $m_{\rho}$  &  $g_{\rm s}^2$  &  $g_{\rm v}^2$  &  $g_{\rho}^2$  
                  &  $\kappa$       &  $\lambda$    &  $\zeta$       &   $\Lambda_{\rm v}$  \\
\hline
\hline
FSUGold2   & 497.479 & 782.500 & 763.000 & 108.094 & 183.789 &  80.466 & 3.0029  & $-$0.000533  & 0.025600  & 0.000823 \\
FSUGarnet  & 496.939  & 782.500 & 763.000 & 110.349 & 187.695 & 192.927 & 3.2600 & $-$0.003551 & 0.023500 & 0.043377 \\
FSUGold2+R   & 501.611 & 782.500 & 763.000 & 103.760 & 169.410 & 128.301 & 3.7924  & $-$0.010635 & 0.011660 & 0.031621  \\
FSUGarnet+R  & 495.633 & 782.500 & 763.000 & 109.130 & 186.481 & 142.966 & 3.2593  & $-$0.003285 & 0.023812 & 0.038274  \\
\hline
\end{tabular}
\caption{Central values for the model parameters FSUGold2 and FSUGarnet before and after Bayesian refinement. The parameter $\kappa$ 
and the meson masses $m_{\rm s}$, $m_{\rm v}$, and $m_{\rho}$ are all given in MeV, and the nucleon mass has been fixed at 
$M\!=\!939$ MeV.}
\label{Table1}
\end{table}
\end{center}
\twocolumngrid


\onecolumngrid
\begin{center}
\begin{figure}[h]
\centering
\includegraphics[width=0.95\textwidth]{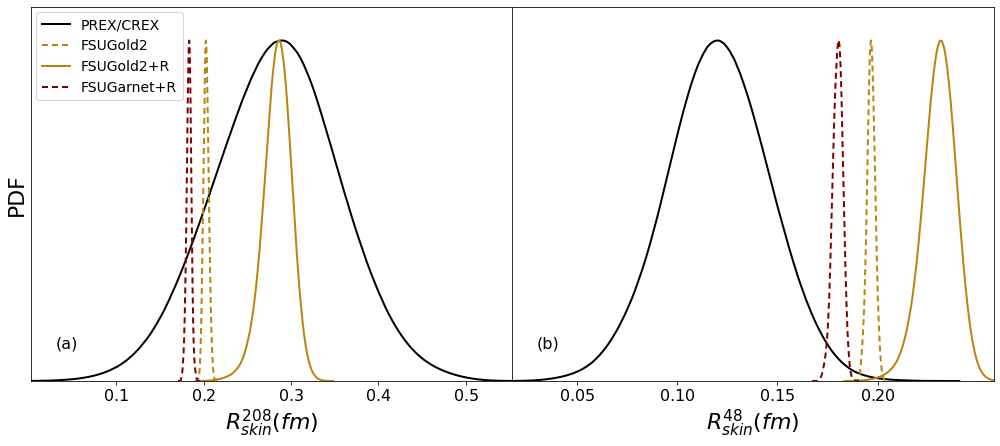}
\caption{Posterior probability distribution functions (un-normalized) for the neutron skin thickness of (a) ${}^{208}$Pb and (b) ${}^{48}$Ca predicted by three of the models listed in Table\,\ref{Table1} are compared against the corresponding experimental result from PREX\,\cite{Adhikari:2021phr} and CREX\,\cite{Adhikari:2022kgg},
	       respectively.} 
\label{Fig2}
\end{figure}
\end{center}
\twocolumngrid

Starting at the lowest rung of the density ladder, we indicated earlier that $\chi$EFT favors a fairly soft EOS 
for pure neutron matter, which according to Eq.(\ref{PvsL}) suggests that the symmetry energy in the vicinity of saturation density is also soft. As we now show, such a softening creates a mild tension when confronted against
the second rung in the density ladder, namely, the parity violating electron scattering experiment on $^{208}$Pb
carried out at Jefferson Lab\,\cite{Adhikari:2021phr}. As already alluded to, the slope of the symmetry energy $L$ 
is highly correlated to the neutron skin thickness of $^{208}$Pb\cite{Brown:2000,Furnstahl:2001un,RocaMaza:2011pm,
Piekarewicz:2019ahf}. The experimental value extracted from PREX and listed in Table\,\ref{Table2}, suggests an
estimate for the slope of the symmetry energy $L\!=\!(106\pm 37)\,{\rm MeV}$\,\cite{Reed:2021nqk} that is significantly 
larger than the $L\!=\!(59.8\pm 4.1)\,{\rm MeV}$ prediction from $\chi$EFT\,\cite{Drischler:2020hwi}. Yet, as indicated
in Table\,\ref{Table2} and Fig \ref{Fig2}, the large value of the neutron skin thickness extracted from PREX is 
in excellent agreement with the predictions from FSUGold2 prior to the refinement. Yet the impact of $\chi$EFT 
on the refinement of the functional is dramatic; the neutron skin thickness of $^{208}$Pb goes down from the  
experimental value of $R_{\rm nskin}^{208}\!=\!0.285\,{\rm fm}$ to $R_{\rm nskin}^{208}\!=\!0.203\,{\rm fm}$.
Note, however, that due to the large experimental uncertainty, such a small value is not yet ruled out. Hence
the need for a more precise determination of $R_{\rm nskin}^{208}$ at the future Mainz facility is well motivated.

In Table\,\ref{Table2} and Fig \ref{Fig2} we also compare our predictions for the neutron skin thickness of $^{48}$Ca 
against the recently completed Calcium Radius EXperiment (CREX)\,\cite{Adhikari:2022kgg}. Although as a medium 
mass nuclei the correlation between $L$ and $R_{\rm nskin}^{48}$ is in general not as strong as for 
$^{208}$Pb\,\cite{Piekarewicz:2012pp}, the class of covariant density functionals used in this work---and indeed most 
theoretical frameworks---suggest a fairly strong correlation between $R_{\rm nskin}^{48}$ and 
$R_{\rm nskin}^{208}$\,\cite{Piekarewicz:2021jte}. Hence, it came as a surprise that $R_{\rm nskin}^{48}$ is 
significantly smaller than $R_{\rm nskin}^{208}$. For this case, the impact of $\chi$EFT moves the FSUGold2 
predictions in the right direction, but not nearly as much as required by CREX. At present, we are not aware of any 
theoretical approach that can simultaneously reproduce both the thick neutron skin in $^{208}$Pb and the thin 
neutron skin in $^{48}$Ca. 

\begin{table}[h]
\begin{center}
\begin{tabular}{|l||c|c|c|c|c|}
 \hline\rule{0pt}{2.25ex}   
 \!\!Model $\left({}^{208}\,{\rm Pb}\right)$ & $R_{\rm ch}$ & $R_{\rm wk}$ & $R_{\rm wskin}$ & $R_{\rm nskin}$ \\
 \hline
 \hline\rule{0pt}{2.25ex} 
 \!\!FSUGold2       &  5.491(6) &  5.801(19) & 0.310(16) & 0.285(15) \\ 
     FSUGold2+R  &  5.517(4) &  5.743(05) & 0.226(03) & 0.203(03)  \\ 
 \hline\rule{0pt}{2.25ex} 
     Experiment   &  5.501(1) &    5.800(75) & 0.299(75) & 0.283(71) \\
 \hline                                                                                                 
 \hline\rule{0pt}{2.25ex}   
 \!\!Model $\left({}^{48}\,{\rm Ca}\right)$ & $R_{\rm ch}$ & $R_{\rm wk}$ & $R_{\rm wskin}$ & $R_{\rm nskin}$ \\
 \hline
 \hline\rule{0pt}{2.25ex} 
 \!\!FSUGold2       &  3.426(3) &  3.707(07) & 0.281(08) & 0.231(08) \\ 
     FSUGold2+R  &  3.477(8) &  3.722(09) & 0.245(02) & 0.197(02)  \\ 
 \hline\rule{0pt}{2.25ex} 
     Experiment     &  3.477(2) &    3.636(35) & 0.159(35) & 0.121(35) \\
 \hline                                                                                                 
\end{tabular}
\caption{Predictions from FSUGold2 before and after refinement (+R) for the charge radius, 
          weak radius, weak skin and neutron skin (all in fm) of ${}^{208}$Pb and ${}^{48}$Ca, 
          as compared with the experimental values extracted from PREX\,\cite{Adhikari:2021phr} 
          and CREX\,\cite{Adhikari:2022kgg}.}
\label{Table2}
\end{center}
\end{table}

Besides constraints from $\chi$EFT that as we just saw have a strong impact on the refinement of the 
functionals, especially in the case of FSUGold2, our previous work also incorporated astrophysical 
constraints on the EOS of neutron star matter from the tidal deformability of GW170817 extracted by 
the LIGO-Virgo collaboration\,\cite{Abbott:PRL2017}, stellar masses and radii of two sources obtained
by the NICER mission\,\cite{Riley:2019yda,Miller:2019cac,Riley:2021pdl,Miller:2021qha}, 
and lower limits on the maximum mass of a neutron star obtained from long time pulsar timing 
observations\,\cite{Cromartie:2019kug,Fonseca:2021wxt}. Indeed, all this information---together with
constraints on the EOS of pure neutron matter predicted by $\chi$EFT---were included in the model
refinement\,\cite{Salinas:2023nci}. By including all this new information in a Bayesian inference
approach, posterior distribution functions were obtained for the neutron star matter EOS, the resulting 
mass-radius relation, and the tidal deformability of a $1.4\,{\rm M}_{\odot}$ neutron star.

\vspace{5pt}
\begin{center}
\begin{figure}[h]
\centering
\includegraphics[width=0.45\textwidth]{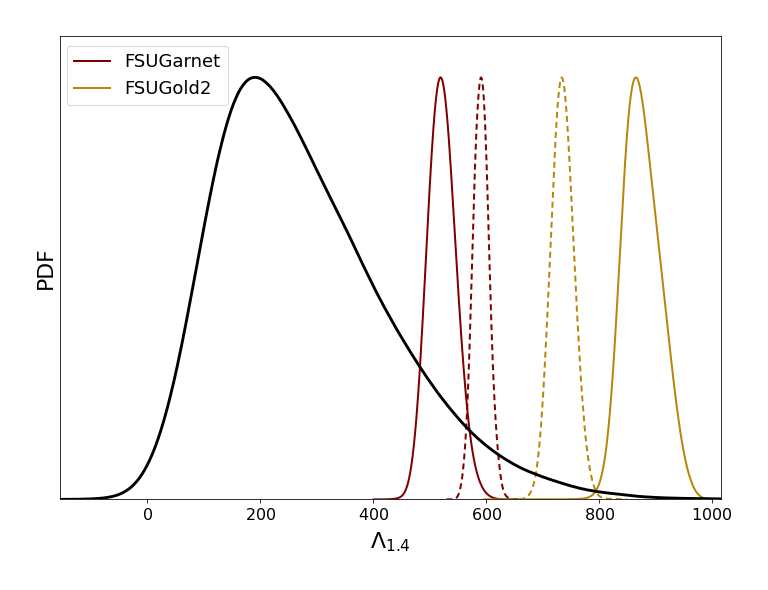}
\caption{Tidal deformability measurement of a $1.4\,{\rm M}_{\odot}$ neutron star\,\cite{Abbott:2018exr} 
along with the predictions from FSUGold2 and FSUGarnet using their respective colors before and after 
refinement; the dashed curves indicate predictions post refinement.}
\label{Fig3}
\end{figure}
\end{center}

Ascending to the fourth rung in the density ladder, we now focus on the gravitational wave profile of
GW170817 that, as we mentioned earlier, allows for the extraction of important structural observables 
such as the chirp mass and tidal deformability. In Fig.\,\ref{Fig3} we examine the model predictions for 
the tidal deformability of a $1.4\,{\rm M}_{\odot}$ neutron star against the recommended value extracted 
by the LIGO-Virgo collaboration of $\Lambda_{1.4} = 190^{+390}_{-120}$, with both the upper and lower 
limits indicating 90\% confidence levels \cite{Abbott:2018exr}. This comparison is quite striking as it 
indicates a significant softening of the EOS at intermediate densities that is not reflected in any of the 
models---even after refinement. If such a discrepancy persists after further scrutiny, see for example 
Ref.\,\cite{Gamba:2020wgg}, this could indicate that the softening may be a reflection of a phase 
transition in the stellar interior.

\begin{center}
\begin{figure}[h]
\centering
\includegraphics[width=0.95\textwidth]{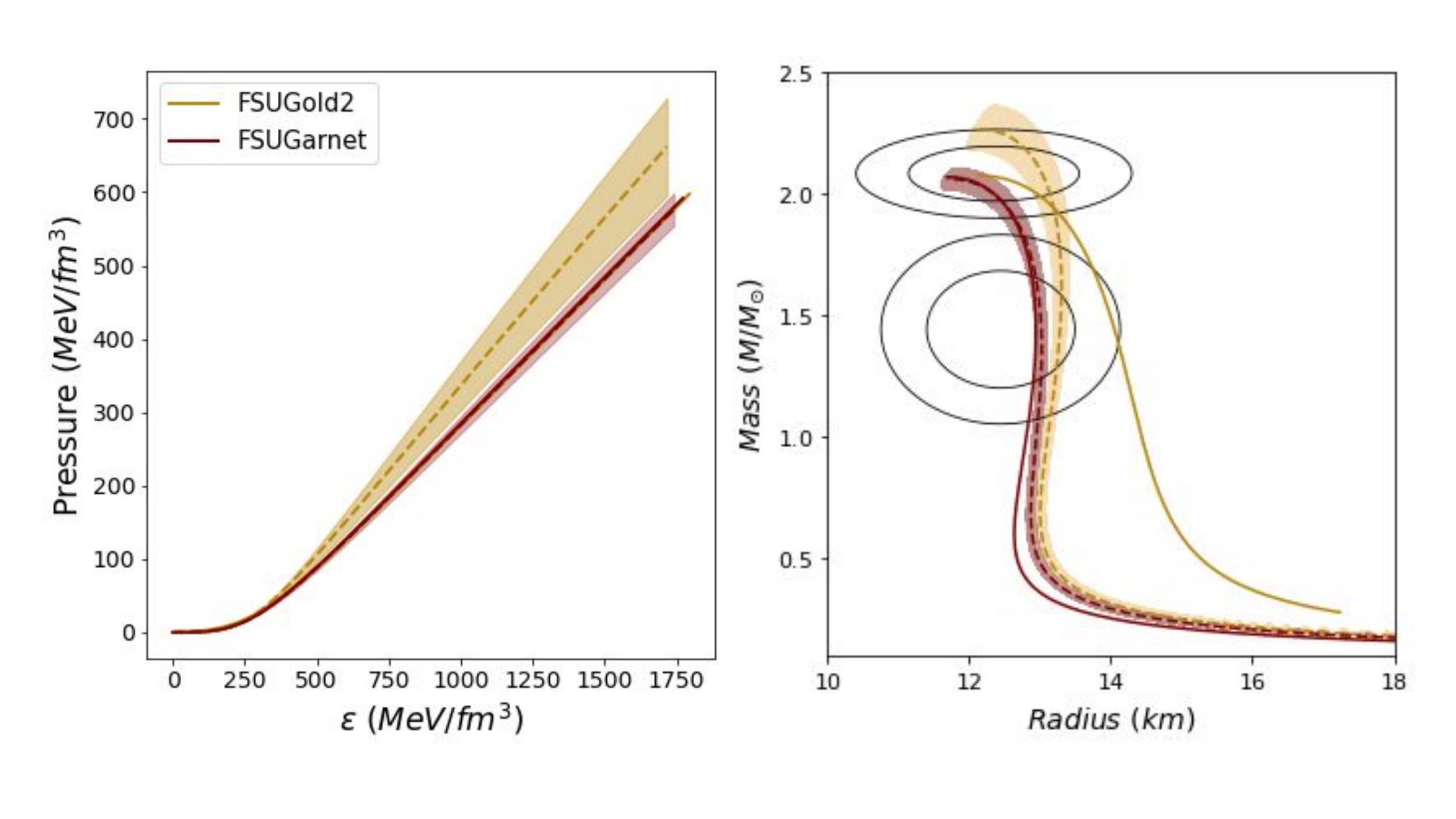}
\caption{(a) Equation of state for neutron star matter as predicted by FSUGold2\,\cite{Chen:2014sca}, FSUGarnet\,\cite{Chen:2014mza}, 
along with the results post refinement\,\cite{Salinas:2023nci}. (b) Mass-Radius relationship predictions from FSUGold2 and FSUGarnet 
are displayed with their respective colors, with the solid and dashed lines representing results before and after the Bayesian refinement, 
respectively. Theoretical error bands for FSUGold2 and FSUGarnet were computed at the 95\% level and the observational covariance 
ellipses represent the 68\% and 95\% confidence intervals.}
\label{Fig4}
\end{figure}
\end{center}

We conclude this section by displaying the holy grail of neutron star structure---the mass-radius relationship---alongside
the neutron star matter equation of state. Recall that there is a one-to-one correspondence between the EOS
and the mass-radius relation\,\cite{Lindblom:1992}, with the EOS providing the microscopic 
underpinning of the macroscopic manifestation. The left-hand panel in Fig.\ref{Fig4} depicts the EOS of neutron
star matter, namely, the EOS of charge-neutral, neutron rich matter in beta equilibrium. Such relation between
the pressure and energy density is the sole ingredient required to solve the Tolman-Oppenheimer-Volkoff 
equations to generate the mass-radius relation. For reference, we note that a value for the energy density of
about $\varepsilon\!\sim\!500\,{\rm MeV}/{\rm fm}^{3}$ correspond to a baryon density of about three
times nuclear matter saturation density. We also note that the refined models are consistent with the limits 
on stellar radii recommended by the NICER mission, depicted in the figure by the 68\% and 95\% 
confidence ellipses. Regardless of whether the softening suggested by the
tidal deformability is confirmed, the EOS at the highest densities found in the core must be stiff enough to
support neutron stars with a mass in excess of two solar masses. Our results indicate that FSUGold2+R is
even consistent with the very large mass of PSR J0952-0607\,\cite{Romani:2022jhd}.

\vspace{5pt}
\onecolumngrid
\twocolumngrid

\section{Conclusions}
\label{Conclusions}

The confluence of major theoretical, experimental, and observational advances in our understanding
of dense matter have motivated the creation of an equation of state density ladder, where the various
rungs in the ladder provide information at specific densities; no single rung can determine the EOS 
over the enormous density range spanned in a neutron star. Moreover, the range of densities probed
by each rung in the ladder overlaps with neighboring rungs, thereby providing consistency checks
among the various methods. Following our recent work\,\cite{Salinas:2023nci} in which previously 
calibrated covariant energy density functionals were refined by the plethora of new information, we
have examined the predictions of the new models. 

First, we concluded that incorporating $\chi$EFT information significantly softens the previously 
stiff FSUGold2 energy density functional. Whereas such a softening shifts the FSUGold2+R 
predictions closer to CREX and LIGO-Virgo, the shift is not nearly as dramatic as the experiment
and observation demand. Moreover, the previous excellent agreement with PREX is now lost. 
Both PREX and CREX will greatly benefit from more precise measurements---which hopefully
may be realized at the future Mainz Energy-recovery Superconducting Accelerator 
(MESA)\,\cite{Becker:2018ggl}. In short, the original FSUGold2 prediction reproduces the PREX 
result, but grossly overestimates the CREX value. As for the FSUGold2+R model, the softening 
of the EOS shifts slightly the result closer to the CREX extraction, but at the cost of losing agreement 
with PREX. This is the best compromise that can be achieved with the present set of covariant energy
density functionals, suggesting that improvements to the isovector sector of the density functional are 
required. Regardless of the limiting set of models used in this contribution, we underscore that
we are not aware of any theoretical approach that can simultaneously reproduce both CREX
and PREX.

Besides $\chi$EFT, the extraction of the tidal deformability of a $1.4\,{\rm M}_{\odot}$ \cite{Abbott:2018exr}
also disfavors a stiff EOS. As shown in this work, all model predictions fall on the high end tail of the
observational value. Given that both NICER and current values for the maximum mass neutron star
require a fairly stiff EOS, if confirmed, the softening suggested by LIGO-Virgo at intermediate densities
may be an indication of a phase transition. At present, such a conclusion is premature given that most
of the observations have large statistical errors. However, given that we have just entered the golden
era of neutron stars, the promise of significant advances in all areas of relevance to the EOS of dense
matter are likely to bring unprecedented precision into the study of neutron stars.

\vfill\eject

\begin{acknowledgments}\vspace{-10pt}
This material is based upon work supported by the U.S. Department of Energy Office of Science, 
Office of Nuclear Physics under Award Number DE-FG02-92ER40750. 
\end{acknowledgments}

\bibliography{./main.bbl}
\end{document}